# Topological superconductivity in lead nanowires

J. G. Rodrigo[1], V. Crespo[1], H. Suderow[1], S. Vieira[1], F. Guinea[2]

[1] Laboratorio de Bajas Temperaturas,
Departamento de Física de la Materia Condensada,
Instituto de Ciencia de Materiales Nicolás Cabrera, Facultad de Ciencias,
Universidad Autónoma de Madrid, E-28049 Madrid, Spain

[2]Instituto de Ciencia de Materiales de Madrid, CSIC
Sor Juana Inés de la Cruz 3
E-28049 Madrid, Spain

Abstract

Superconductors with an odd number of bands crossing the Fermi energy have topologically protected Andreev states at interfaces, including Majorana states in one dimensional geometries. Superconductivity, a low number of 1D channels, large spin orbit coupling, and a sizeable Zeeman energy, are present in lead nanowires produced by nanoindentation of a Pb tip on a Pb substrate, in magnetic fields higher than the Pb bulk critical field. A number of such devices have been analyzed. In some of them, the dependence of the critical current on magnetic field, and the Multiple Andreev Reflections observed at finite voltages, are compatible with the existence of topological superconductivity.

PACs. 74.25.-q, 73.63.-b, 73.63.Nm

A number of materials have band structures which support edge and surface states with unusual charge and spin transport properties[1-3]. These materials include generalized Integer Quantum Hall systems, topological insulators, and topological superconductors. The excitations at the edges of one dimensional topological superconductors can be described as Majorana particles[4-6]. The exchange of two such states leads to a non trivial modification of the state of the system. The simplest realization of a topological superconducting state requires[4-13] i) a small number of conduction channels, ii) a band structure modified by spin-orbit coupling, iii) an interaction which leads to the formation of Cooper pairs, and iv) a sufficiently strong Zeeman coupling to an external magnetic field. The ingredients described above are present in lead nanowires a few angstroms wide in the presence of a magnetic field higher than the bulk critical field[14-19].

The system studied here is sketched in Fig.1. A narrow and elongated constriction between two lead electrodes is built by carefully stretching an STM tip away from a substrate. Close to the breaking point, the number of conducting channels is small and their characteristics can be obtained from the Multiple Andreev Reflection (MAR) spectra[20-23]. The superconducting properties of the electrodes and the constriction are modified in an applied magnetic field. The system continues to exhibit a Josephson current at zero voltage and MAR peaks at fields larger than the bulk critical field, $H_c$. At these fields the electrodes are in the normal state, and superconductivity is restricted to the constriction, where orbital currents cannot quench superconductivity. The resulting device can be seen as a nanoscopic Josephson junction, with a weak link where the voltage drop occurs. The magnetic field also induces a Zeeman splitting on the electrons in the constriction.

As the two electrodes are unequal, the magnetic field is more effective in changing the superconducting features in one of them, which eventually becomes normal. When this happens, the superconducting gap is lowered, and a significant Zeeman shift of the bands can be expected. Spin-orbit coupling in lead is large, and the estimated *g* factors for bulk lead are in the range $g \approx 4-6$ [24,25], which can be enhanced by interaction effects in nanoscopic samples [26]. For fields in the range $H \approx 0.1 - 0.2\, T$, the Zeeman splitting, $B$, can be of order of $0.04 - 0.06\, meV$, while the superconducting gap, $\Delta = 1.35\, meV$ at zero field and zero temperature, is expected to go smoothly to zero as the magnetic field increases. Hence, a regime where the Zeeman coupling is larger than the superconducting gap can exist in some of the samples studied here. Topological superconductivity requires that the Fermi energy lies within the Zeeman gap induced by the magnetic field. The position of the Fermi energy at the constriction depends on details of the electrostatic potential which, in turn, is determined by the geometry of the contact. We expect that this is the case in a fraction of the samples studied, due to random fluctuations in the electrostatic potential.

When the right combination of parameters is achieved, the Zeeman coupling will open a gap near the Fermi level, so that the number of pairs of bands crossing the Fermi energy will be odd on one side of the constriction. These are the conditions required for topological superconductivity to exist. The constriction becomes a boundary between a topological and a non topological superconductor, S-$S_T$. A midgap state with particle-hole character will be formed there. Another broad resonance with mixed particle-hole character is expected at the N-$S_T$ junction where the superconducting features disappear away from the constriction. The two resonances will be hybridized and changed into conventional Andreev states when they are closer than the superconducting coherence length. If the coupling between the two states can be neglected, the midgap state at the constriction has all the features of a Majorana particle. The regions S and $S_T$ have random rough edges. They are in the diffusive regime, with an elastic mean free path, $\ell$, comparable to their width *W*. Hence, a Majorana fermion at the constraint will be well defined if the length of the $S_T$ region, *L*, is such that $L \geq \sqrt{\xi \ell} \approx \sqrt{\xi W}$, where $\xi \approx 80$ nm is the coherence length in clean lead.

In the experiments, indentations of a Pb tip on a Pb sample are induced, in order to fabricate sharp elongated nanotips and nano-protrusions on the sample surface. The experiments on the resulting nanostructures are performed at 0.3 K, with the STM installed in a ³He cryostat equipped with a superconducting solenoid to apply a magnetic field. The evolution of the electronic and superconducting properties of the nanostructure versus magnetic field can be followed from the analysis of the conductance characteristics of the constriction. During the magnetic field sweeps, the STM feedback loop is kept active, at fixed bias voltage, with a constant value for the current across the constriction, in order to ensure that the overall geometry of the nanostructure is not altered along the process. The feedback loop is blocked during the acquisition of the current vs. voltage curves. Different nanostructures, with conductance values at the constriction ranging from 2 $G_0$ to 50 $G_0$ ($G_0$ is the conductance quantum) were studied. The numeric derivative of the I-V curves acquired during the magnetic field sweeps gives the conductance, where the signatures of the different Andreev Reflection processes can be easily identified. The presence of Josephson current, a finite current at zero bias, reflects a sharp peak in the conductance curves at zero bias. The I-V characteristics in the absence of Zeeman coupling shows distinct features at $V = \Delta + \Delta'$ and at $V = \Delta + \Delta'$, where $\Delta$ and $\Delta'$ are the superconducting gaps at the two regions at each side of the junction.

In order to investigate the phenomenon described above, a nanocontact with low conductance is created at zero magnetic field, and its electronic and superconducting properties will be followed as a function of the magnetic field. We focus on the variations with field of the conductance of the junction, the value of the Josephson critical current, and the detail of the Andreev Reflection features present in the conductance curves, as shown in fig. 2.

The I-V curves obtained at zero field are fitted to the MAR model to obtain the number of conducting channels and its transmission values. Following the procedure described in[27] we get that four channels, with transmission values 1, 0.920, 0.600 and 0.225 account for 99.5% of the current, being the contribution of other channels below 0.005, which is the limit of the resolution in the fitting. This result indicates that the condition requiring a small number of conducting channels to observe Majorana particles is fulfilled. During the sweep of the magnetic field the STM feedback loop is active in order to keep the situation of the contact as stable as possible. Nevertheless slight atomic rearrangements at the nanocontact may take place. These rearrangements, which reflect as small jumps and variations of the conductance of the contact (fig. 2(a)), may lead to large changes in the value of the Josephson critical current, $I_C$ (fig. 2(b)). This is a consequence of the variations of the individual transmissions of the channels involved in the contact, and how the electromagnetic environment affects the phase coherence required to have Josephson current for different values of the channel transmission[28,29]. Therefore, it is possible to find situations in which two contacts, presenting the same current at a given finite voltage, have a different conductance, and the one with lower conductance presents the highest critical current.

The effect of these rearrangements can be detected during the sweep of the magnetic field up to its bulk critical value (75 mT at 300 mK), while the conductance curves keep similar MAR features (curves *a-b* in fig.2(c)). The crossing of $H_c$ is detected in the experiment by the onset of a progressive reduction (in voltage and intensity) of the MAR feature at high bias, a sharp decrease of the Josephson critical current (to about half of the average value below $H_c$), and an upturn of the conductance. This upturn can be related to changes in the excess current as the magnetic field is reducing more effectively the superconducting features in one of the nanoelectrodes.

As field is increased above $H_c$ the MAR features and the Josephson critical current are progressively reduced until 125 mT, where we detect an unexpected rise of the critical current, with a maximum at 150 mT and a continuous decrease at higher fields. This is accompanied by the evolution of the Andreev Reflections signature in the conductance curves towards a SN situation, but with a well defined Josephson-like signature at zero bias (curves *d-f* in fig. 2(c)) up to 4 $H_c$.

We checked the robustness of this observation by repeating the field sweeps. The "anomalous" bump in the evolution of the Josephson current at high field was observed several times, until in one of the sweeps the abovementioned atomic rearrangements led to the situation presented in fig. 3. After these rearrangements, at about 40 mT, the nanocontact presented higher conductance but a clearly smaller Josephson critical current. The characteristics of the conducting channels before and after the rearrangements were obtained as above, and we find that in the new configuration up to eight channels contribute with transmission values above 0.1, being less than 0.4 for five of them [28].

As the field is further increased we obtain the usually expected evolution of the conductance curves and the Josephson critical current. There is a sharp jump in $I_C$ at $B_c$ followed by a progressive reduction of the MAR signature and the value of $I_C$, until 130 mT where the conductance curves present a SN type Andreev Reflection behavior, and no Josephson-like feature can be detected at zero bias. This evolution, presented in fig. 3, is practically identical to the "standard" results obtained for larger nanocontacts, with conductances in the range of 50 $G_0$ and above [28].

We have modeled the above results by generalizing MAR scattering theory to a partially open channel which connects a topological, $S_T$, and a non topological superconductor, S [28,30]. Typical examples are shown in Fig.[4]. The high voltage structure is washed out as the Zeeman coupling increases, and a single feature at about the value of the highest superconducting gap remains for Zeeman couplings near and above the transition. As the magnetic field is increased, the dependence of the Josephson current on the transmission coefficient evolves from $I_c \propto T$ in the S-S' regime, to $I_c \propto \sqrt{T}$ in the $S_T$-$S'_T$ regime[31] (for $T \ll 1$) leading to a minimum in $I_c$ in the S-$S'_T$ regime. The suppression of structure in the I-V curves at high voltages, and the minimum in the value of the critical current can be explained by the existence of a junction between a non topological and topological superconductor.

The results presented here suggest that narrow lead constructions are a good system where to realize boundaries between topological and non topological superconductors (see also [32]). Midgap states which give rise to Majorana fermions can exist at these boundaries. The nano-constrictions studied here show simultaneously superconductivity, few channels, strong spin-orbit coupling, and a large modification of the superconducting features by a magnetic field. These junctions can be fabricated in large numbers, and it is expected that, in some of them, the different interactions have the right values for the existence of Majorana fermions. These systems are an interesting alternative to other materials currently under study in the search for Majorana fermions in condensed matter physics[33,34].

The Laboratorio de Bajas Temperaturas is associated to the ICMM of the CSIC. This work was supported by the Spanish MINECO (Consolider Ingenio Molecular Nanoscience CSD2007-00010 program, FIS2011-23488, ACI-2009-0905), by the Comunidad de Madrid through program Nanobiomagnet. F. G. acknowledges funding from grants FIS2008-00124, FIS2011-23713 and from the ERC Advanced Grants program, contract 290846. This research was supported in part by the National Science Foundation under Grant No. NSF PHY11-25915. F. G. acknowledges useful conversations with E. Prada, P. San Jose, and B. Trauzettel.

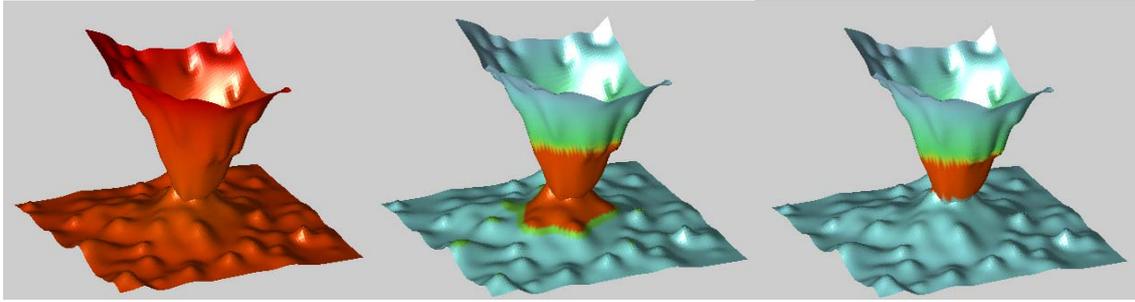

Fig.1 Sketch of the nanostructure and nano-electrodes involved in the experiments. The superconducting region is colored red. Left: At low applied magnetic fields, $H \ll H_c$, where $H_c$ is the bulk critical field of lead, the whole structure is in the superconducting state. Center: For $H \geq H_c$ superconductivity is restricted to the region near the junction, and the device shows a finite Josephson current. Right: For $H \gg H_c$ superconductivity has disappeared in one electrode, and the Josephson current vanishes. At still higher fields the whole device is in the normal state.

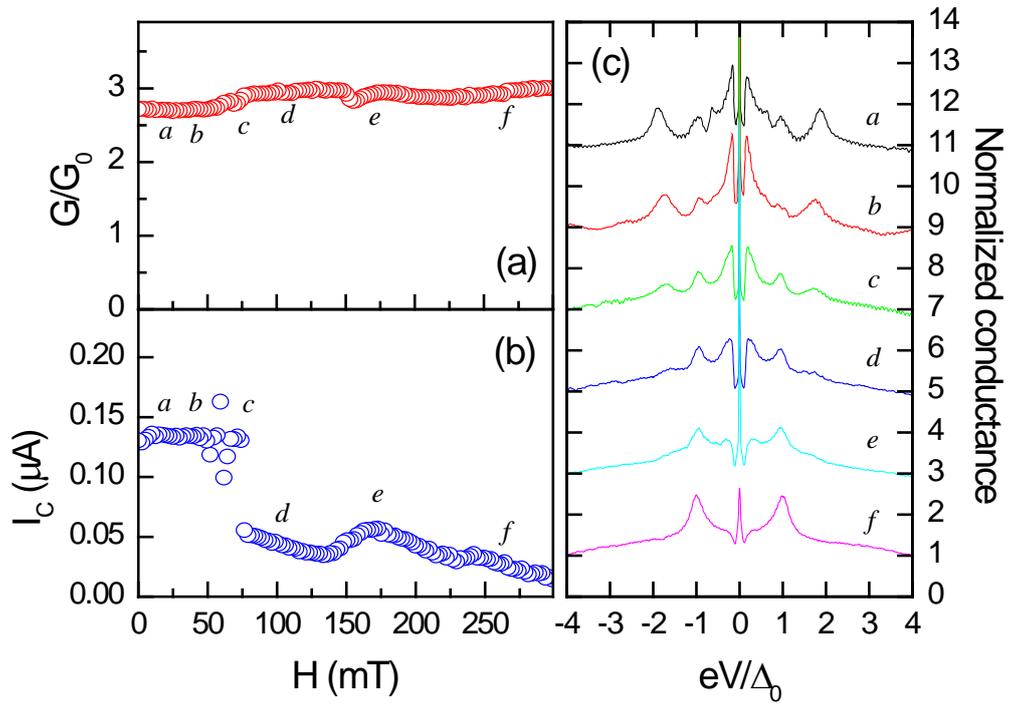

Fig.2 Evolution of the conductance, (a), and Josephson current, (b), of a narrow constriction ($G \approx 3G_0$) as a function of the applied magnetic field. Note the bump of the Josephson current at $\approx 175\ mT$. In (c) we present several conductance curves obtained along the field sweep. The field values corresponding to the curves are indicated with the labels *a-f* in panels (a) and (b). ($\Delta_0$ is the value of the superconducting gap of lead at zero field, 1.35 meV. Curves are shifted vertically 2 units for clarity).

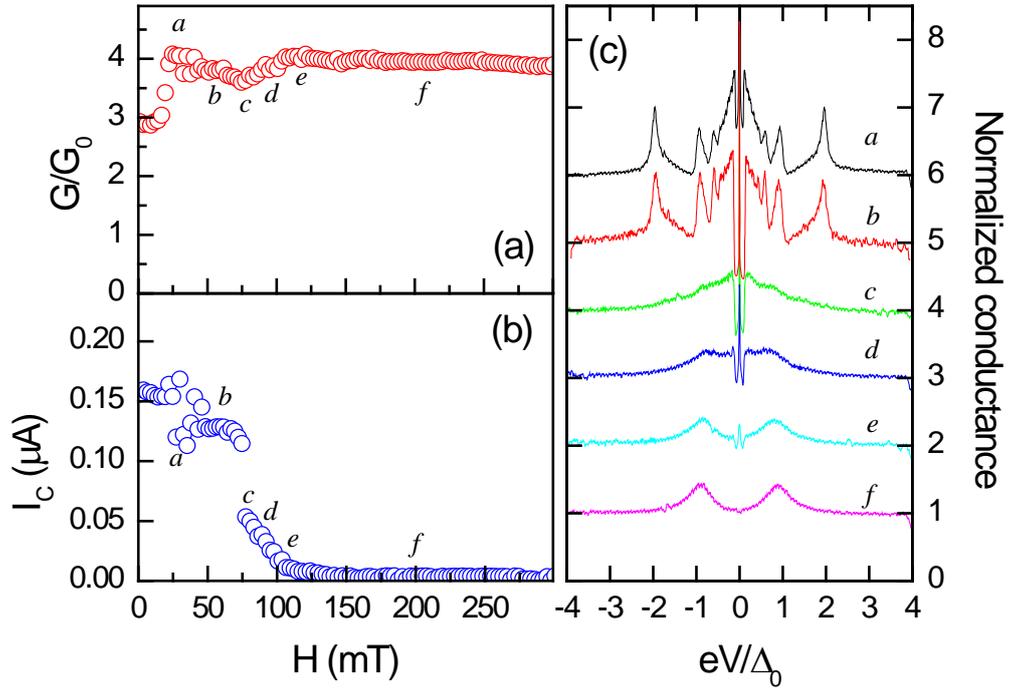

Fig.3 Evolution of the conductance, (a), and Josephson current, (b), of a narrow constriction ($G \approx 4G_0$) as a function of the applied magnetic field. In (c) we present several conductance curves obtained along the field sweep. The field values corresponding to the curves are indicated with the labels *a-f* in panels (a) and (b). ($\Delta_0$ is the value of the superconducting gap of lead at zero field, 1.35 meV. Curves are shifted vertically 1 unit for clarity).

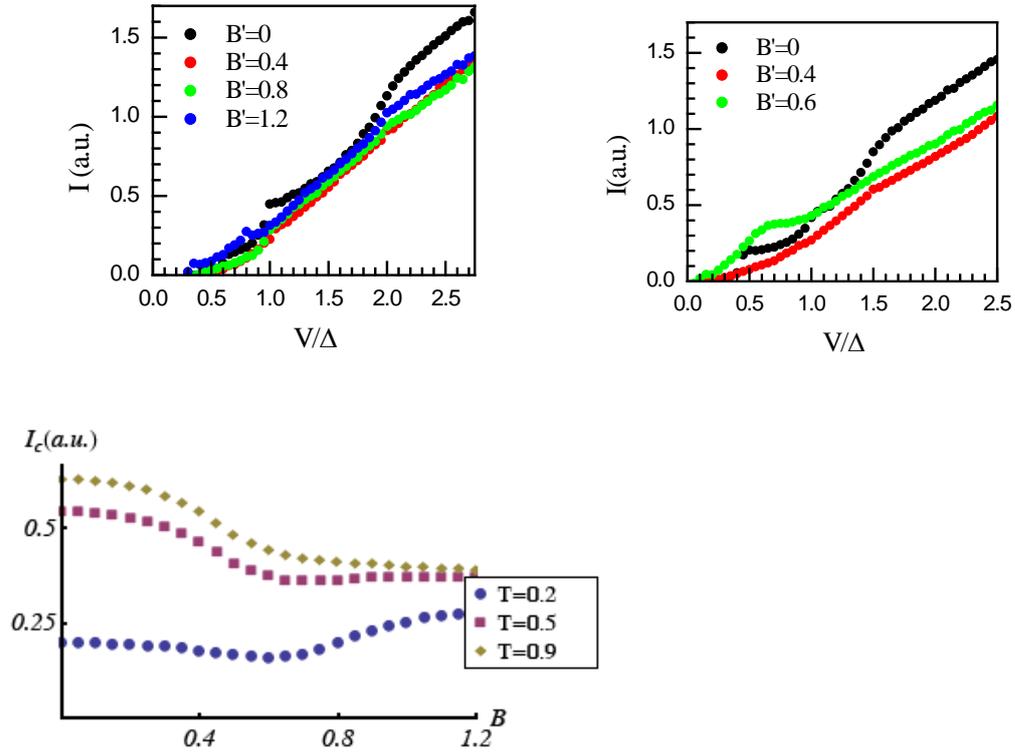

FIG. 4: Calculated I-V curves of junctions with different values of the Zeeman coupling. (a): $\Delta=\Delta'=1$. (b): $\Delta=1, \Delta'=0.5$. (c): Critical current as function of Zeeman coupling for a superconducting junction with $\Delta = 0.8$, $\Delta' = 0.5$ and different transmissions (see Supplementary Information). The junction type is S-S' for $0 \leq B \leq 0.4$, S-$S_T$ for $0.4 \leq B \leq 0.8$, and $S_T$-$S'_T$ for $0.8 \leq B$

# Supplementary information

**Multiple Andreev reflection at junctions between topological and non topological superconductors.**

We analyze the Multiple Andreev Scattering (MAR) at the interface between a topological and a non topological superconductor using the scattering approach in[S1].

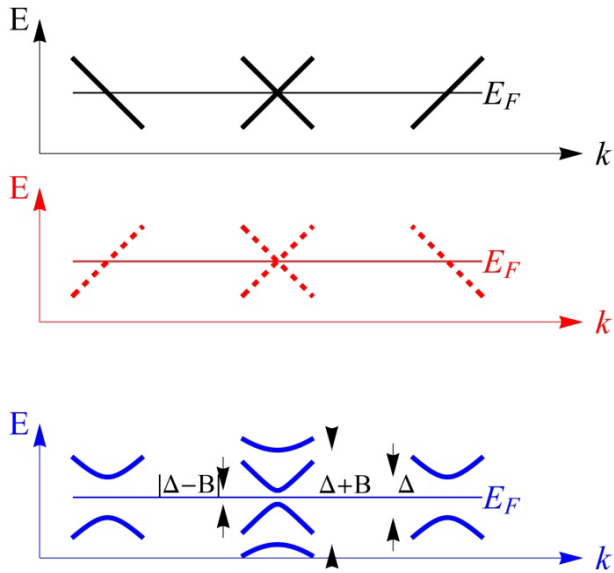

Fig. S1. Electronic structure of the channel discussed in the text. Top: Spin up and spin down electron bands in the absence of superconducting and Zeeman gaps. Center: hole bands. Bottom: Electronic structure for finite superconducting and Zeeman gaps.

We consider a single channel. A finite spin-orbit coupling separates the spin up and spins down bands. The Zeeman coupling induced by an applied magnetic field mixes different spins, and, finally, the superconducting gap hybridizes electron and hole states. For simplicity, we assume that the Fermi energy lies at the intersection of the two central bands. The resulting electronic structure is sketched in Fig.[S1].

We define the junction by a spin independent scattering matrix, which mixes right and left propagating states at each side of the junction, as in[S1]. Each spin channel defines two $2 \times 2$ matrices, one for electrons, and one for holes. The final scattering matrix has dimensions $8 \times 8$, divided into four $2 \times 2$ boxes.

The combined effect of the superconducting and Zeeman gaps change the scattering states. We assume that the superconducting and Zeeman gaps are different at the two electrodes, $\Delta, \Delta', B, B'$. The Hamiltonian which describes a given electrode is:

$$H = \begin{pmatrix} vk & 0 & 0 & 0 & 0 & 0 & 0 & \Delta \\ 0 & -vk & 0 & 0 & 0 & 0 & -\Delta & 0 \\ 0 & 0 & vk & B & 0 & -\Delta & 0 & 0 \\ 0 & 0 & B & -vk & \Delta & 0 & 0 & 0 \\ 0 & 0 & 0 & \Delta & vk & -B & 0 & 0 \\ 0 & 0 & -\Delta & 0 & -B & -vk & 0 & 0 \\ 0 & -\Delta & 0 & 0 & 0 & 0 & vk & 0 \\ \Delta & 0 & 0 & 0 & 0 & 0 & 0 & -vk \end{pmatrix} \qquad \text{S1}$$

where we assume that $\Delta_i$ is real, $v$ is the Fermi velocity, and $k$ is the momentum. The diagonalization of the Hamiltonian gives eight scattering states, defined by the values of $\Delta$, $B$, and the energy, $E = vk$. An electron (hole) injected from the left with energy $E$ can be reflected as an electron (hole) with the same energy, or it can be transmitted as an electron (hole) with energy $E \pm V$. After many reflections, the initial incoming electron spawns scattering states with energies $E + 2nV$ on the left side, and $E + (2n + 1)V$ on the right electrode, with $n = -\infty, \ldots, \infty$. The scattering states of the Hamiltonian can be expressed using the matrix:

$$A(E,\Delta) = \begin{pmatrix} \alpha(E,\Delta) & 0 & 0 & \alpha^{-1}(E,\Delta) & 0 & 0 & 0 & 0 \\ 0 & -\alpha^{-1}(E,\Delta) & -\alpha^{-1}(E,\Delta) & 0 & 0 & 0 & 0 & 0 \\ 0 & 0 & 0 & 0 & -\alpha(E,\Delta+B) & -\alpha^{-1}(E,\Delta+B) & -\alpha(E,\Delta-B) & -\alpha^{-1}(E,\Delta-B) \\ 0 & 0 & 0 & 0 & -1 & -1 & 1 & 1 \\ 0 & 0 & 0 & 0 & -\alpha(E,\Delta+B) & -\alpha^{-1}(E,\Delta+B) & \alpha(E,\Delta-B) & \alpha^{-1}(E,\Delta-B) \\ 0 & 0 & 0 & 0 & 1 & 1 & 1 & 1 \\ 0 & 1 & 1 & 0 & 0 & 0 & 0 & 0 \\ 1 & 0 & 0 & 1 & 0 & 0 & 0 & 0 \end{pmatrix} \qquad \text{S2}$$

where:

$$\alpha(E,\Delta) = \begin{cases} \dfrac{E + i\sqrt{\Delta^2 - E^2}}{\Delta} & E \leq \Delta \\[2mm] \dfrac{-E - \text{Sign}(E)\sqrt{E^2 - \Delta^2}}{\Delta} & E > \Delta \end{cases} \qquad \text{S3}$$

The scattering states can be grouped into four sets on the left electrode, and four sets on the right electrode, connected by the scattering matrix. We define the associated amplitudes as $c_{k,n}$ where $k = 2,4,6,8$ denote amplitudes in one electrode, and $k = 1,3,5,7$ denote amplitudes in the other electrode. The equations to be solved are

$$\begin{pmatrix} r & 0 & 0 & t \\ 0 & -r & t & 0 \\ 0 & t & r & 0 \\ t & 0 & 0 & -r \end{pmatrix} \begin{pmatrix} \sum_{k=2,4,6,8} A_{k,1}(E+2nV,\Delta,B)c_{k,n} + \delta_{0,n}\delta_{i,1}J \\ \sum_{k=1,3,5,7} A_{k,2}[E+(2n+1)V,\Delta',B']c_{k,n} \\ \sum_{k=2,4,6,8} A_{k,3}(E+2nV,\Delta,B)c_{k,n} + \delta_{0,n}\delta_{i,2}J \\ \sum_{k=1,3,5,7} A_{k,4}[E+(2n+1)V,\Delta',B']c_{k,n} \end{pmatrix} = \begin{pmatrix} \sum_{k=2,4,6,8} A_{k,4}(E+2nV,\Delta,B)c_{k,n} \\ \sum_{k=1,3,5,7} A_{k,3}[E+(2n+1)V,\Delta',B']c_{k,n} \\ \sum_{k=2,4,6,8} A_{k,2}(E+2nV,\Delta,B)c_{k,n} \\ \sum_{k=1,3,5,7} A_{k,1}[E+(2n+1)V,\Delta',B']c_{k,n} \end{pmatrix}$$

$$\begin{pmatrix} r & 0 & 0 & t \\ 0 & -r & t & 0 \\ 0 & t & r & 0 \\ t & 0 & 0 & -r \end{pmatrix} \begin{pmatrix} \sum_{k=2,4,6,8} A_{k,5}(E+2nV,\Delta,B)c_{k,n} \\ \sum_{k=1,3,5,7} A_{k,6}[E+(2n-1)V,\Delta',B']c_{k,n-1} \\ \sum_{k=2,4,6,8} A_{k,7}(E+2nV,\Delta,B)c_{k,n} \\ \sum_{k=1,3,5,7} A_{k,8}[E+(2n-1)V,\Delta',B']c_{k,n-1} \end{pmatrix} = \begin{pmatrix} \sum_{k=2,4,6,8} A_{k,8}(E+2nV,\Delta,B)c_{k,n} \\ \sum_{k=1,3,5,7} A_{k,7}[E+(2n-1)V,\Delta',B']c_{k,n-1} \\ \sum_{k=2,4,6,8} A_{k,6}(E+2nV,\Delta,B)c_{k,n} \\ \sum_{k=1,3,5,7} A_{k,5}[E+(2n-1)V,\Delta',B']c_{k,n-1} \end{pmatrix} \quad \text{S4}$$

Where the index $i = 1,2$ stands for the channel in which current is injected. For simplicity, we assume that the transmission and reflection amplitudes, $t$ and $r$, are real.

In addition to a calculation for each injecting channel, the equations need to be solved when the current comes from the other electrode, as the two electrodes are not assumed to be equivalent. This implies the replacement $1 \leftrightarrow 2$ in the superconducting and Zeeman gap indices. The total number of amplitudes which need to be computed is twice the number needed in the analysis of ordinary MAR[S1], and the number of scattering equations to be solved is multiplied by four. The same degree of additional complexity applies to the calculations performed for a single spin polarized channel carried out in [S2].

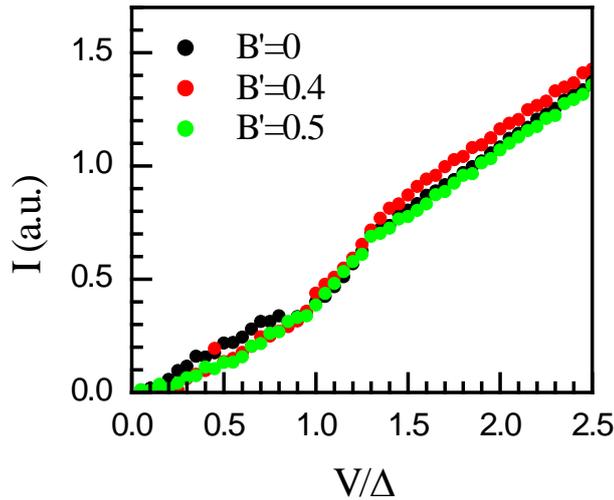

Fig. S2 Calculated I-V characteristics for a junction between inequivalent superconductors with different values of the Zeeman coupling $B'$ (in all cases, $\Delta = 1$, $\Delta' = 0.3$ and $B = 0$). The transmission coefficient is $\tau = 0.5$.

**Calculation of the critical current.**

We calculate the critical current of a junction between two s-wave superconductors with spin-orbit and Zeeman coupling by using a discrete version of the Hamiltonian, and minimizing the total energy as function of the phase difference across the junction. A superconducting channel is written as the sum of four chains, which stand for the two spin species, and electrons and holes. The Hamiltonian of one electrode is

$$H = H_R + H_L + H_J$$
$$H_{R,L} = H_{R,L}^{kin} + H_{R,L}^{so} + H_{R,L}^{Z} + H_{R,L}^{\Delta}$$
$$H_{R,L}^{kin} = t \sum_{n=\pm 1,\ldots \pm \infty} \left( c_{n\uparrow}^{+} c_{n+1\uparrow} + c_{n\downarrow}^{+} c_{n+1\downarrow} \right) + h.c.$$
$$H_{R,L}^{so} = i\alpha \sum_{n=\pm 1,\ldots \pm \infty} \left( c_{n\uparrow}^{+} c_{n+1\uparrow} - c_{n\downarrow}^{+} c_{n+1\downarrow} \right) + h.c. \qquad \text{S5}$$
$$H_{R,L}^{Z} = B \sum_{n=\pm 1,\ldots \pm \infty} c_{n\uparrow}^{+} c_{n\downarrow} + h.c.$$
$$H_{R,L}^{\Delta} = \Delta \sum_{n=\pm 1,\ldots \pm \infty} c_{n\uparrow}^{+} c_{n\downarrow}^{+} + h.c.$$
$$H_J = t' \left( c_{-1\uparrow}^{+} c_{1\uparrow} + c_{-1\downarrow}^{+} c_{1\downarrow} \right) + h.c.$$

We choose $\varepsilon_F = -2t$ so that the chemical potential coincides with the crossing of the spin up and spin down bands, as shown in Fig.[S1]. We fix the transmission coefficient $\tau$, which determines the value of $t'$

$$\frac{t'}{t} = \sqrt{1 + \frac{\alpha^2}{t^2} \frac{(2-\tau)}{T} - \frac{2}{\tau}\left(\frac{\alpha}{t}\right)\sqrt{(1-\tau)\left(\tau + \frac{\alpha^2}{t^2}\right)}} \qquad \text{S6}$$

Without loss of generality, we set $t = 1$. The problem is defined by the values of $\alpha, \tau, \Delta_1, \Delta_2, B, B'$ and the phase $\phi$ across the junction. The critical current, $I_c$ is defined by the maximum value of $|\partial E(\phi)/\partial \phi|$. Plots of $E(\phi)$ and $\partial E(\phi)/\partial \phi$ and $\tau = 0.5$ are shown in Fig. S3. The calculations are limited by the maximum number of sites, which set a lower bound on the allowed superconducting and Zeeman gaps. The features of a topological superconductor are reasonably described, in a lattice with 200 sites, by $\alpha/t = \sqrt{3}$. This choice of parameters implies that the gaps in the system are not much lower than the total bandwidth, defined as the distance between the Fermi energy and the bottom of the band.

Results for the dependence of the total energy on the superconducting phase difference, $E(\phi)$, are shown in Fig. S3, for the three transmission coefficients, $T = 0.2, 0.5, 0.9$ used in Fig. 4c) of the main text. The critical current for $T \ll 1$ evolves from $I_c \propto T$ at low fields, where the junction is S-S', to $I_c \propto \sqrt{T}$ in the $S_T$-S'$_T$ regime[S4] at large fields, leading to a minimum in $I_c$ at intermediate fields. In the $S_T$-S'$_T$ regime there is an Andreev state at zero energy for $\phi = \pi$ and the value of $E'(\pi)$ is different from zero.

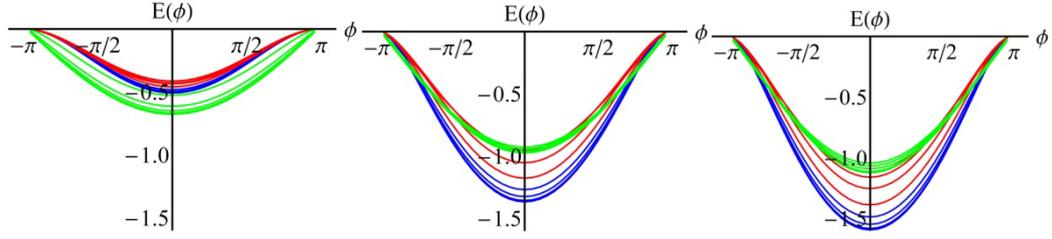

Fig. S3. Dependence of the energy of the junction on the superconducting phase difference, $E(\phi)$, for $\Delta = 0.8, \Delta' = 0.4$ and different magnetic fields, as in Fig. 4c) of the main text. Left: $T = 0.2$, center, $T = 0.5$, right, $T = 0.9$. The magnetic fields lie in the range $0 \leq B \leq 1.6$. The blue curves describe an S-S' junction, $0 \leq B \leq 0.4$, the red curves describe an S-S$_T$ junction, $\leq B \leq 0.8$, and the green curves describe an S$_T$-S'$_T$ junction, $0.8 \leq B$.

**Evolution of a broad nanocontact under magnetic field.**

In fig. S4 we present the evolution with magnetic field of the conductance of the contact, the value of the Josephson critical current, and the detail of the Andreev Reflection features present in the conductance curves, for a broad constriction ($G = 46G_0$) formed between a Pb tip and a Pb sample. As the two electrodes are unequal, the magnetic field is more effective in changing the superconducting features in one of them, which eventually becomes normal. For magnetic field below $H_c$, (75 mT for Pb at 300 mK) the conductance curves present the expected MAR features, with peaks at $2\Delta_0/n$ (n=1,2,3,...), as well as a sharp peak at zero bias corresponding to Josephson current (see I-V curves also). Between 75 mT and 130 mT there is a strong reduction of the peak at $2\Delta_0$, becoming just a slight bump, and it is accompanied by a reduction of the Josephson current (see $I_c$ vs H plot). This corresponds to a progressive destruction of superconductivity in the "weaker" nano-electrode. Finally, above 130 mT we obtain the standard NS Andreev conductance curves, with no signature of Josephson-like current, as only one nano-electrode remains superconducting.

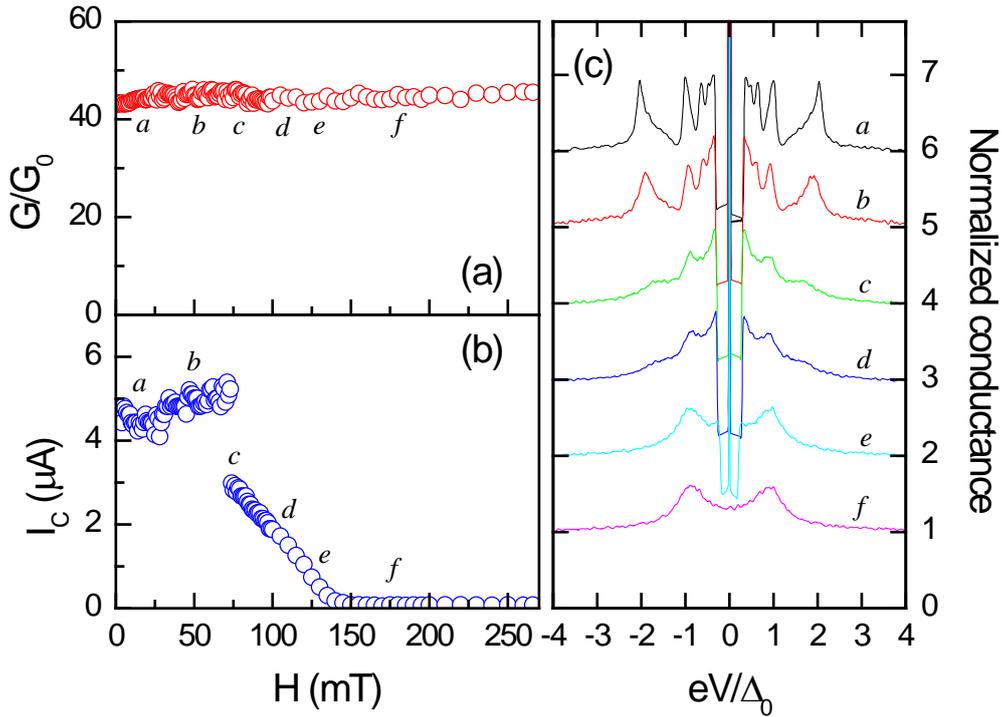

Fig.S4 Evolution of the conductance, (a), and Josephson current, (b), of a broad constriction ($G \approx 46G_0$) as a function of the applied magnetic field. In (c) we present several conductance curves obtained along the field sweep. The field values corresponding to the curves are indicated with the labels *a-f* in panels (a) and (b). ). ($\Delta_0$ is the value of the superconducting gap of lead at zero field, 1.35 meV. Curves are shifted vertically 1 unit for clarity).

**Conducting channels in the nanocontacts.**

In fig. S5 we present the analysis of the quantum conducting channels involved in the different nanocontacts studied in this work. The analysis is done in terms of MAR, following the procedure described in ref [S3]. Despite the difference in the number of channels involved, note the similarity between the results obtained for the constrictions corresponding to the cases shown in fig.3 and fig. S4, regarding the ratio between the number of channels with high transmission (i.e., >0.5) and low (<0.5), and its contribution to the total conductance. Their difference, compared to the results for the contact in fig.2 [S5.a], should be considered in order to account for the observability of Majorana fermions in this type of nanostructures.

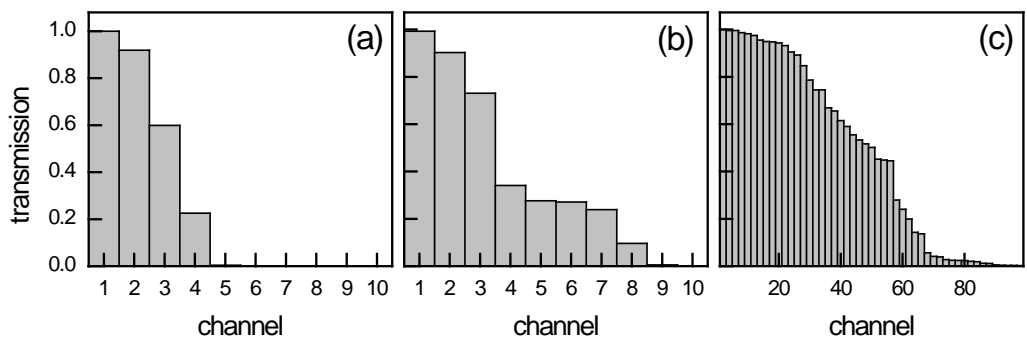

Fig.S5  (a) Fit of channel transmissions for the junction whose MAR curves are shown in Fig.2. (b) Fit for the junction in Fig. 3. (b) Fit for the junction in Fig. S4.